\documentclass[conference]{IEEEtran}
\IEEEoverridecommandlockouts
\usepackage{cite}
\usepackage{amsmath,amssymb,amsfonts}
\usepackage{algorithmic}
\usepackage{graphicx}
\usepackage{textcomp}
\usepackage{xcolor}
\def\BibTeX{{\rm B\kern-.05em{\sc i\kern-.025em b}\kern-.08em
    T\kern-.1667em\lower.7ex\hbox{E}\kern-.125emX}}

\usepackage{xspace}
\newcommand{\Meta}{Meta\xspace}

\newcommand{\Bento}{Bento\xspace}

\newcommand{\FBDetect}{FbDetect\xspace}

\newcommand{\ServiceLab}{ServiceLab\xspace}

\newcommand{\AdsManager}{Ads Manager\xspace}

\newcommand{\blindedref}{~\cite{valdez2018real}\xspace}

\newcommand{\blindedrefii}{~\cite{adsmanager}\xspace}

\usepackage{amsmath}
\usepackage{algorithmic}
\usepackage{graphicx}
\usepackage{textcomp}
\usepackage{enumitem}
\usepackage{bbold}
\usepackage{comment}
\usepackage[utf8]{inputenc}
\usepackage{graphicx}
\usepackage{hyperref}
\usepackage{verbatim}
\usepackage{cleveref}
\usepackage{listings}
\usepackage{booktabs}
\usepackage{subfigure}
\usepackage{framed}
\usepackage{balance}
\usepackage{pifont}

\usepackage{xcolor}
\definecolor{LightGray}{gray}{0.9}
\usepackage[frozencache=true,cachedir=minted-cache]{minted}

\graphicspath{{./figs/}}

\begin{document}
\title{Learning to Learn to Predict Performance Regressions in Production at \Meta}







\author{\IEEEauthorblockN{Moritz Beller,\IEEEauthorrefmark{1} Hongyu Li,\IEEEauthorrefmark{2}\\ Vivek Nair,\IEEEauthorrefmark{1} Vijayaraghavan Murali,\IEEEauthorrefmark{1} Imad Ahmad,\IEEEauthorrefmark{1} Jürgen Cito,\IEEEauthorrefmark{2}\IEEEauthorrefmark{3} Drew Carlson,\IEEEauthorrefmark{2} Ari Aye,\IEEEauthorrefmark{1} Wes Dyer\IEEEauthorrefmark{1}}
\IEEEauthorblockA{
\IEEEauthorrefmark{1}\textit{Meta Platforms, Inc., Menlo Park, USA}, \, \IEEEauthorrefmark{2}\textit{Ex-Meta}, \, \IEEEauthorrefmark{3}\textit{TU Wien, Austria,} \\
\{mmb, viveknair, vijaymurali, imadahmad, gaa, wesdyer\}@meta.com, \, juergen.cito@tuwien.ac.at}
}

\maketitle

\begin{abstract}


Catching and attributing code change-induced performance regressions in production is hard; predicting them beforehand, even harder.
A primer on automatically learning to predict performance regressions in software, this article gives an account of the experiences we gained when researching and deploying an ML-based regression prediction pipeline at \Meta.

In this paper, we report on a comparative study with four ML models of increasing complexity, from (1)~code-opaque, over (2)~Bag of Words, (3)~off-the-shelve Transformer-based, to (4)~a bespoke Transformer-based model, coined SuperPerforator. Our investigation shows the inherent difficulty of the performance prediction problem, which is characterized by a large imbalance of benign onto regressing changes. Our results also call into question the general applicability of Transformer-based architectures for performance prediction: an off-the-shelve CodeBERT-based approach had surprisingly poor performance; even the highly customized SuperPerforator architecture achieved offline results that were on par with simpler Bag of Words models; it only started to significantly outperform it for down-stream use cases in an online setting. To gain further insight into SuperPerforator, we explored it via a series of experiments computing counterfactual explanations. These highlight which parts of a code change the model deems important, thereby validating it.

The ability of SuperPerforator to transfer to an application with few learning examples afforded an opportunity to deploy it in practice at \Meta: it can act as a pre-filter to sort out changes that are unlikely to introduce a regression, truncating the space of changes to search a regression in by up to 43\%, a 45x improvement over a random baseline.

\end{abstract}

\section{Introduction}
A performance regression of a software system describes a situation in which the system  exhibits correct functional properties, but does so using significantly more resources than before the introduction of the regression~\cite{luo2016mining}. Often, performance regressions are user-facing, \emph{e.g.}, the system becomes slow to respond. This has the potential to cause significant business impact. Performance has thus been described as ``one of the most influential non-functional requirements of software''~\cite{1291833}, ``with the power to make or break a software system in today's competitive market''~\cite{6224281}.

For software developers, performance regressions often pose a challenge and can be harder to fix than correctness bugs~\cite{shang:15,zaman2011security}, since the conditions under which the regressions occur are often intricate and the result of a complex interplay of configuration and code changes. Moreover, testing for correctness is typically a binary problem, whereas assessing acceptable values of performance involves a gray zone, making it an ambiguous problem by nature.

Similar to other defects~\cite{planning2002economic}, it is beneficial to detect performance regressions as early as possible in the software development life cycle. At an early stage, (1) fewer users are affected, (2) root causes can more easily be localized and fixed, (3) no changes have been built on top of the regression-inducing change, which might need to be reverted, and (4) one averts the computational cost of running a sub-optimal code base. However, from our experience at \Meta, we know that even the complex safety net of modern linters~\cite{harmim19scalable, tricorder:15}, a state-of-the-art static complexity bounds checker~\cite{DBLP:conf/sas/CicekBCD20}, and a comprehensive performance regression test suite are unable to catch all regressions before they land in production.

For this reason, \Meta employs two more regression detection systems: \ServiceLab, which performs A-B experiments by replaying past live traffic on suspect changes, and a production monitoring system called \FBDetect. However, neither is without drawbacks: \ServiceLab does not have the bandwidth and would be a bottleneck if all code changes triggered an experiment on it; \FBDetect only finds the damage after it has been done. Moreover, sometimes in distributed systems the size and scale of {\Meta}'s, regressions can go unnoticed by \FBDetect, since there is a complex, interwoven architecture of systems such as load balancers, dynamic infrastructure scalers, and caches that conceal and buffer the effects of increased CPU utilization, excessive memory consumption, or latency~\blindedref. In such a system, even seemingly insignificant regressions can accumulate over time.

For this reason, we want to enable a transformative shift towards earlier detection of performance regressions in the development workflow at \Meta, outside of the traditional and existing tools. Such a new technique must be (1) fast to apply, (2) highly scalable, and (3) accurate in its predictions.

In this paper, we introduce SuperPerforator, a pre-trained, doubly-fine-tuned, Transformer-based deep neural net architecture to accurately predict the risk of a performance regression given a code change: SuperPerforator is first pre-trained on a very large corpus of millions of code changes at \Meta, then fine-tuned on hundreds of thousands of precise, real-world production performance measurements from \FBDetect. We then employ a final transfer learning step to fine-tune it a second time on a handful of performance regressions in a different domain for its final use case with \Meta.

Our results highlight the inherent difficulty of the problem: conducting accurate performance measurements as the training base for the Machine Learning (ML) algorithms in the right format and quantity is a challenge by itself; any automated learning approach is plagued by the very large imbalance of benign onto performance-inducing changes, ranging at a ratio of lower than 100:1 at \Meta; simpler models employing traditional ML algorithms such as Random Forests and operating on non-code features might be helpful in describing the conditions that lead to the introduction of performance regression in the past, but are not nearly sensitive enough to predict their occurrence in the future. Off-the-shelve application of highly popular Deep Learning (DL) based models perform equally poorly. Surprisingly, a Bag of Words model achieved offline results that were on par with a sophisticated, highly customized Transformer-based architecture tailored toward the problem, coined SuperPerforator. These results call into question the general applicability of DL-based and  Transformer-based models in particular for the performance regression prediction task.
However, the generalization and transferability of SuperPerforator afforded an opportunity for us to deploy it in practice at \Meta, and it showed 10x improved performance over the BoW in this live setting.

We conclude our investigation with a validation of SuperPerforator by producing counterfactual explanations to understand which code parts it focuses on and a number of open questions and challenges for future research.

In short, this paper makes the following contributions:
\begin{itemize}
    \item A comparative study across ML techniques of various degree of refinement to predict performance regressions
    \item The introduction of SuperPerforator, an ML architecture to predict performance regressions in production
    \item A deep investigation into the workings of SuperPerforator, including counterfactual-based explanations
\end{itemize}

\section{Background}

\subsection{Literature}

There has been a plethora of literature that discusses performance analysis and modeling techniques of software systems~\cite{cortellessa2011model,balsamo2003simulation,xu2012rule,trubiani2011detection,leitner2017exploratory}, using a multitude of different viewpoints and approaches to the problem.

Traditionally, performance models in literature are designed to capture operational metrics (\emph{e.g.}, CPU, latency, memory consumption, I/O writes) in relation to an underlying software system and a workload. They can be used to predict performance properties of software systems for different purposes. These include workload characterization and capacity planning~\cite{workload1, workload2, workload3}, configuration selection~\cite{config1, config2, config3}, and anomaly detection~\cite{anomaly1, anomaly2}. Our work is different in that it is concerned with the \emph{prediction} of performance regressions based on \emph{source code changes}. We therefore focus here on the relevant literature on how performance regressions have been detected through black-box ML models, of which there is a surprising shortage. Such ML and other statistical modeling techniques models typically do not incorporate deep \emph{a priori} knowledge about the system’s internal behavior, but are rather constructed by training a classifier on a corpus of previously detected performance regressions.

Shang et al.~\cite{shang:15} propose an approach that builds a statistical model to detect a regression between two software versions by first clustering performance counters based on correlation analysis. Similarly, AutoPerf~\cite{alam:19} collects hardware performance counters of two consecutive versions of a program by running performance regression test cases. They train an auto-encoder using the collected profiles of the previous version and run it against the profiles of the next version. They then classify run instances as performance regressions if the reconstruction error of the auto-encoder exceeds a certain threshold. From a technical perspective, AutoPerf uses unsupervised learning, while we use  supervised approaches trained on many code change pairs by solely considering static features (such as the code change itself), \emph{i.e.,} without the need to collect profiling information.

In summary, unlike our approach, these approaches require dedicated test runs while we learn a model that explicitly avoids the need to run such computationally expensive tests.

The work by Liao et al.~\cite{liao:20} tries to discern whether, given two versions of the system and two different workloads, observable differences in response times can be attributed to the new workload or new version of the system. Like our work, this paper describes attempts to train different kinds of models. However, they re-use the same features across models, whereas we experiment with different features and other kinds of input spaces as well as with different model architectures.

Perhaps closest in nature and concept to our work is Ithemal~\cite{mendis2019ithemal}, which tries to predict the number of clock cycles a processor takes to execute a block of assembly instructions. It shows that an LSTM architecture can outperform the prediction accuracy for throughput measurements from static rules. Essentially, we want to do the same with SuperPerforator, but several abstraction levels higher working on larger pieces of source code changes written in high-level languages instead of the very limited set of op codes.



\subsection{Tools and processes at \Meta}
\subsubsection{DiffBERT}
\label{sec:diffbert}

\begin{figure}[tb]
\centering
\includegraphics[width=\columnwidth]{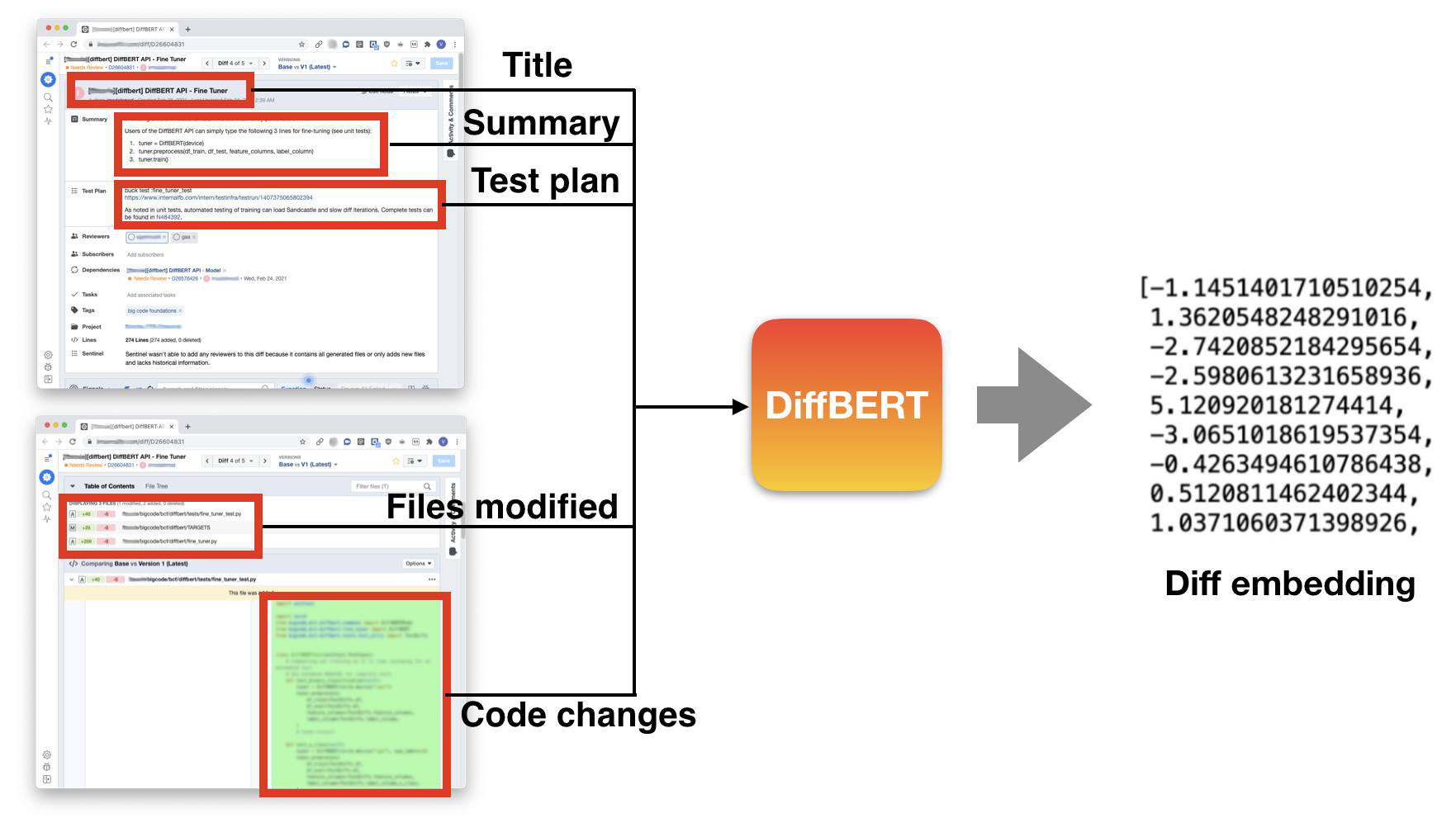}
\caption{DiffBERT embedding process in action.}
\label{fig:diffbert}
\end{figure}

DiffBERT is a pre-trained multi-modal and multi-lingual DL model that can embed code changes in dense vector space. It improves upon CodeBERT in several aspects and is programmatically built on top of the Huggingface Python library~\cite{huggingface, feng2020codebert}. DiffBERT is available as an API at \Meta.

At its core, DiffBERT takes a diff and outputs an encoding, as shown in \Cref{fig:diffbert}. A diff at \Meta (left side of \Cref{fig:diffbert}) represents a peer-reviewed code change together with a title, summary, test plan, and the list of modified files.
These features encompass different modalities, comprising natural language, code, and file paths, all of which DiffBERT can ingest. In addition, DiffBERT is also designed to work across language barriers,
including Hack---an advanced PHP dialect~\cite{Hack}---,
Java, Python, JavaScript, HTML, and others, since a typical diff at \Meta may simultaneously change code in multiple languages.

To incorporate this multi-modality and multi-linguality, DiffBERT had to be able to work with very large input documents. Particularly, transformer-based models like BERT suffer from the bottleneck of self-attention~\cite{wies2021transformer}, a mechanism that is quadratic in the model's input length. The Longformer introduces a notion of local self-attention, which trades off a small amount of accuracy for the ability to scale to much larger inputs~\cite{beltagy2020longformer}; we therefore adapted this model for DiffBERT. DiffBERT is pre-trained with the Masked Language Model (MLM) objective that is used in models like CodeBERT. In this self-supervised objective, given a sequence of input tokens, a random 15\% of the tokens are masked, and the model is trained to predict them back using signals from the rest. By doing this, the model learns to produce a good representation (embedding) for a given input.

SuperPerforator uses DiffBERT as the underlying model but makes two specific changes: (1) it removes the natural language features like title and summary (since these are not available when working on code), and (2) it does a dedicated pre-training on the 18 million code changes that are subject to \FBDetect measurement (instead of on all changes at \Meta).


\subsubsection{\FBDetect}
\label{sec:fbdetect}

\FBDetect is {\Meta}'s regression detection and management system. It continuously monitors performance metrics in production, represented by time series, for non-anticipated level-shifts. When a shift is detected, \FBDetect determines its potential root causes and creates alerts. These alerts are tracked until resolution. Concretely, \FBDetect can measure runtime-based regressions expressed as the CPU percentage the regressed function takes up over the total amount of CPU utilization across the entire \Meta fleet, called \emph{GCPU}.

\FBDetect measures CPU utilization in its server fleet through a sample-based heuristic: every few seconds, it selects a random batch of servers. It then freeze-frames each selected server and records the current function execution stack. This calculation of function performance is inclusive of how much time is spent in the function itself plus all its upstream, calling functions. This sample-based measurement strategy has the by-product that functions which execute very quickly or are executed very rarely are less likely to appear in the measurement table. For this reason, the majority of (at a company level) insignificant functions never show up in the performance statistics, or if they do, then by chance. Another corollary is that seldom invoked functions might exhibit high GCPU variance due to the small amount of sampled data.

\subsubsection{\AdsManager}
\label{sec:adsmanager}

\begin{figure}[tb]
\centering
\includegraphics[width=\columnwidth]{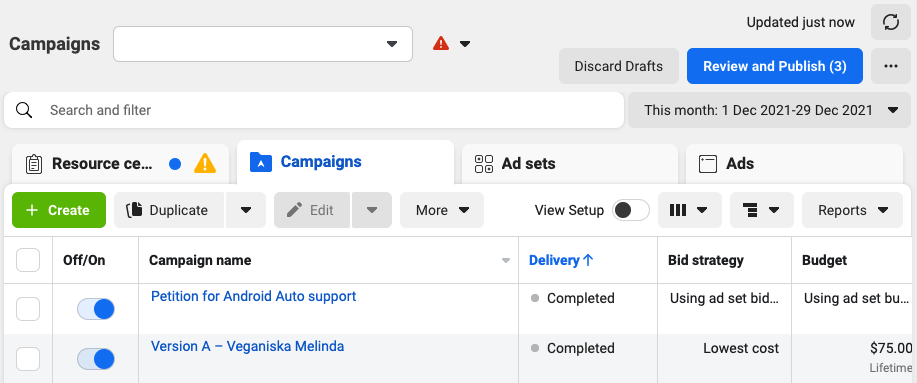}
\caption{\AdsManager.}
\label{fig:am}
\end{figure}

Millions of advertisers use \AdsManager, depicted in \Cref{fig:am}, on a daily basis to purchase, manage, and analyze their advertising campaigns. It allows advertisers to edit and organize ad campaigns, analyze their ad performance, distribute budgets, or run split tests~\blindedrefii.

As a consequence of its frequent use, performance regressions in \AdsManager can have severe business impact and \AdsManager already has a number of tools to detect regressions before they land in production. However, these existing tools are computationally expensive and therefore can not operate on all potential diffs. \AdsManager needed a solution that would reliably trim the search space of diffs for their downstream regression detection tools for both frontend and backend code.

\subsubsection{{\Bento}.}
\label{sec:bento}
All our experiments, unless otherwise stated, were performed on \Bento notebooks (a {\Meta}-internal version of Jupyter~\cite{kluyver2016jupyter}), using on-demand servers equipped with either 2xP--100, 2xV--100, or 8xP--100 nvidia GPUs, a 20 core Intel Xeon Gold 6138 CPU @ 2.00GHz, and 256G RAM.

\section{Research Design}
In this paper, we perform a comparative study of four approaches to model performance regressions, in increasing order of refinement: (1) Code opaque models looking at diff-level features that do not take into account the code (2) Bag-of-words models in combination with Random Forest classifiers that do consider code, but not its structure (3) Representative of off-the-shelf DL models, a pre-trained version of CodeBERT with no modifications, and (4) a \Meta-bespoke, heavily pre-trained and fine-tuned Transformer-based model called SuperPerforator.

We evaluate models purely by their prediction capabilities, as opposed to their ability to fit to seen data. We try to give all models fair chances by controlling for equal operating conditions as much as possible (\emph{e.g.}, we fix a certain temporal train-test split). From an effort perspective, we roughly increased person hours put into each of the models by an order of magnitude.

Finally, we explore the role of how much code context provides the best prediction results and investigate properties of the SuperPerforator model and why it makes certain predictions through counterfactual explanations.

\subsection{Data}
\label{sec:thedata}

\begin{figure*}[tb]
\centering
\includegraphics[width=1.35\columnwidth]{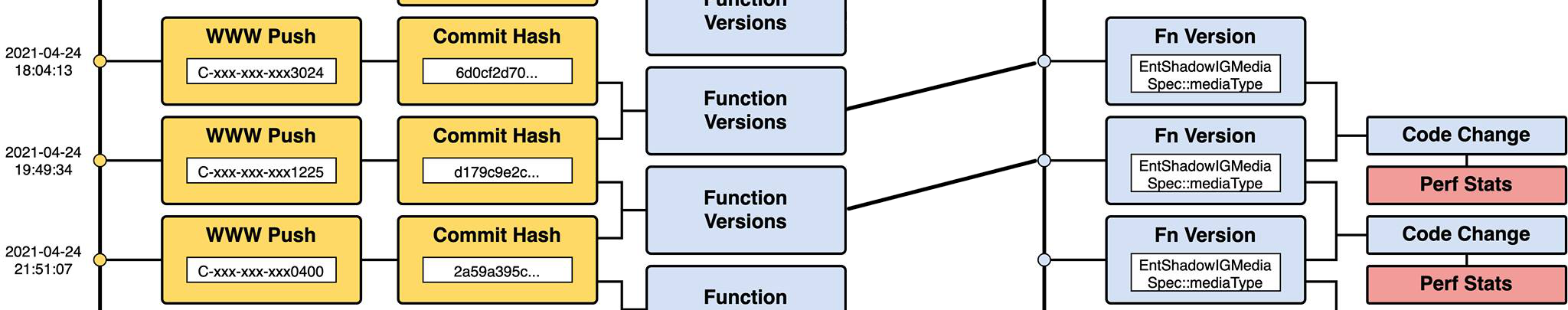}
\caption{Acquiring function-level performance changes.}
\vspace{-0.2cm}
\label{fig:funlevel}
\end{figure*}

For SuperPerforator, we use {\FBDetect}'s regression detection and combine it with source control and release cycle data to obtain fine-grained function performance change metrics of the type {\tt <time stamp $t_n$, fully qualified function name, function source code change, $\Delta$CPU utilization>}. This way, we can associate a code change to a function (or, more precisely, the union of function changes rolled out as one release, typically on an hourly cadence) with its impact on CPU performance, given there is a significant impact on performance, as detected by \FBDetect.

\Cref{fig:funlevel} depicts this process in detail. From each release (``www push'') on the left-hand side, we extract the latest commit hash associated with it and identify which functions changed in the release by comparing its commit to the preceding releases' commit. To generate the final database, for each modified function version identified in commit comparison, we look up its full function content before and after the release, as well as its performance behavior before and after. If there is no code change in a function for subsequent www pushes, we would merge this time frame into the previous tuple, since that code change ``lives on.'' In practice, this means that some function code change tuples can have performance stats over many weeks, while others are short-lived.  Thus, we assemble a database of not only regressions, but also improvements to function performance, and their associated code changes.

Finally, precisely predicting by \emph{how much} a function regressed in CPU utilization turned out to be both too difficult and unnecessary; merely identifying \emph{that} there is a significant regression provides a cleaner signal for developers. We hence converted the more difficult regression problem of predicting the exact regressed value into a binary classification task, where we classify a function change as either inducing a regression or not.

\subsection{Function stability}


As mentioned in \Cref{sec:fbdetect}, some of the measured functions can show unstable behavior, due not to code changes, but to the sampling measurement process itself. It is paramount to detect and remove such outliers, as  learning algorithms tend to be particularly upset by the presence of noisy binary training data~\cite{natarajan2013learning}.

Another feature of \FBDetect measurements is that its CPU utilization measure $GCPU$ is absolute---if a function has no substantial change to it, but starts to be called more often, then its CPU utilization in the fleet will increase. {\FBDetect}'s idea behind this is that a slow function that is called rarely is not problematic at company scale.

\subsection{Descriptive statistics}

For the experiments in this paper, unless otherwise stated, we consumed diffs
written in Hack
from {\Meta}'s www repository over a period of 6 months, from June 1st to November, 30th, 2021.
This leads to a data set of 484,285 function version pairs, and 457,226 after removing invalid data points (\emph{e.g.}, data that had an empty change associated with it, which might stem from pipeline errors). Out of these, 3,390 (0.7\%) represent performance regressions, highlighting the very large imbalance in the data set.

Coefficient of variance (CV, \cite{hopkins1997new}) is a standardized measure of dispersion of a probability distribution calculated as $\frac{\sigma}{\mu}$.
Finally, we binarize $\Delta_{GCPU_t} = GCPU_t - GCPU_{t-1}$ into regressions and non-regressions, by running a simplified version of {\FBDetect}'s regression detection logic that essentially marks a code change as a regression if $\Delta_{GCPU_t}$ exceeds a threshold $t$.

\subsection{Chronological train and test split}
\label{sec:chrono}
To ensure that training does not learn from the future, we separate train and test sets chronologically.

In early experiments, we found clusters of diffs that are all marked as regressions that are not only chronologically but also syntactically close; if we did not separate them by time and naïvely applied a random 80:20 split, the model would learn the properties of the cluster in the train set, making it trivial to predict results for the other diffs from the same cluster in the test set. Our results for such a non-chronological split show at least two times better prediction performance metrics $f_1$. Consequently, we define the entire month of November as our test set, and the five months before it as our train set.

\subsection{Code context}
\label{sec:codecontext}

Diffs at \Meta are usually represented with the default code context setting of the Unix {\tt diff} utility, that is, one line of code context above and below each code chunk in the diff. For example, Listing~\ref{lst:inconspicuous}, generated with the default context length, looks relatively inconspicuous at first. However, given more code context, its potential to regress becomes obvious in Listing~\ref{lst:more_context}.

\begin{listing}[bt]

\begin{minted}
[
frame=lines,
framesep=2mm,
baselinestretch=1.2,
bgcolor=LightGray,
fontsize=\footnotesize,
escapeinside=@@,
startinline
]{python}
    print(i)
+   call_medium_expensive_function(i)
\end{minted}
\caption{Seemingly inconspicuous code change.}
\label{lst:inconspicuous}
\vspace{-0.5cm}
\end{listing}

\begin{listing}[bt]
\begin{minted}
[
frame=lines,
framesep=2mm,
baselinestretch=1.2,
bgcolor=LightGray,
fontsize=\footnotesize,
escapeinside=@@,
startinline
]{python}
for i in (1, all_users):
    print(i)
+   call_medium_expensive_function(i)
\end{minted}
\caption{More context reveals the potential for a regression.}
\label{lst:more_context}
\vspace{-0.5cm}
\end{listing}

We then hypothesized that the surrounding context might play a more significant role for the detection of performance regressions. Of course, there is a trade-off between having so much surrounding context that it might drown the signal from the actual change, and running into model limitations, (\emph{e.g.,} for DiffBERT-based models, the maximum document length of its Longformer).
For this reason, we performed a series of experiments on various code context lengths to determine the optimal value empirically.

\section{Code-opaque model}
Our first attempt to model performance prediction was to use code-opaque features, \emph{i.e.}, features that do not involve a direct representation of the source code.

The aim for this model was not to optimize its performance in the single decimal percentages, but to gauge how well it would generally perform. We chose this general approach initially to set a reasonable baseline with traditional ML that did not require a lot of model engineering.

\subsection{Method}
For this model we used author-related features team name, tenure class at \Meta (expressed in 7 different quanta) and change-related features such as the number of changed files and the number of changed source lines of code~\cite{albrecht1983software}, the base file paths they changed (truncated at maximum 3 folder segments deep to avoid over-specification), as well as the extensions of the changed files. To minimize the risk of overfitting and reduce dimensional complexity of our training data, we filtered the categorical inputs to assign entries a default category with fewer than five occurrences (for team, file paths, and extensions). We then multi-hot-encoded the file path and file extensions, one-hot encoded the other categorical features, and left the remaining features as-is.

Finally, we ran a Gradient Boosting Classifier using 100 estimators, a learning rate of 0.1 and a maximum depth of 3, and a Random Forest Classifier using 1,000 estimators and balanced class weights to cater for the imbalance in the labels.

\subsection{Results}
\label{sec:code-results}

\begin{figure*}[tbh]
\centering
\includegraphics[width=0.6\columnwidth]{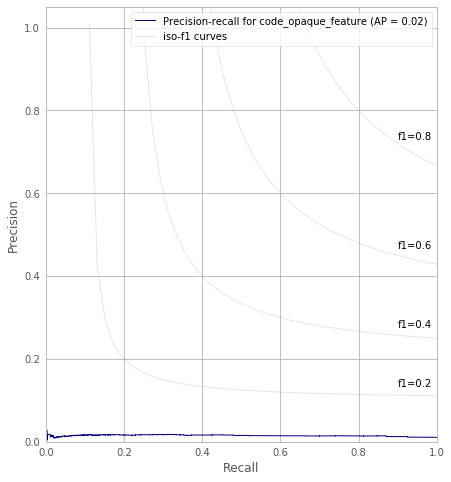}
\includegraphics[width=0.6\columnwidth]{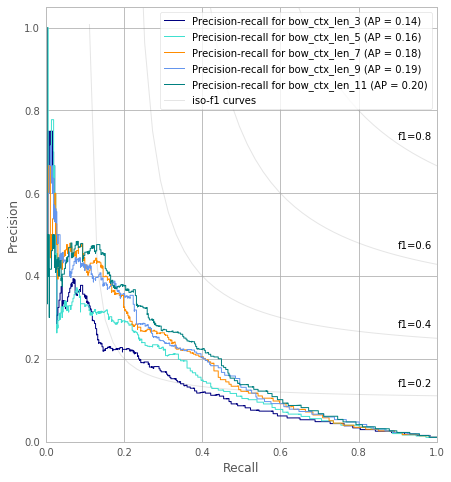}
\includegraphics[width=0.6
\columnwidth]{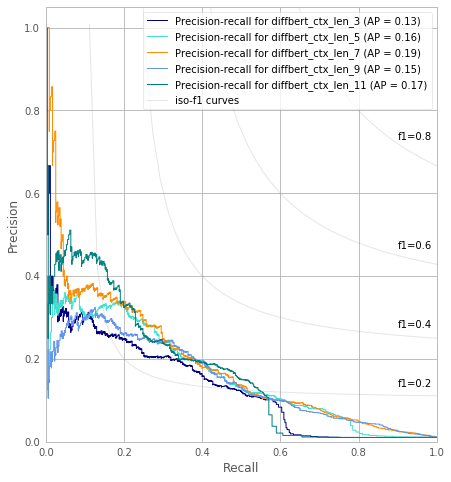}
\caption{ROC AUC~\cite{narkhede2018understanding} and precision recall curves of (1) Code-opaque feature  (Random Forest), (2) BoW (random forest), and (3) DiffBERT models. Omitted CodeBERT's graph since its flat line is identical to sub plot (1).}
\label{fig:bow_results}
\end{figure*}

\Cref{fig:bow_results} depicts precision-recall curves for the different models. With the curve of the first subplot showing essentially no lift from the y axis, results are underwhelming in terms of their \emph{predictive capabilities} for the code-opaque model. As the large data imbalance might be responsible for this, we experimented with oversampling and automatic oversampling approaches such as SMOTE~\cite{chawla2002smote}, but found that they only marginally improved results.

At the same time, the model fit of the code opaque feature model on the train set was relatively high (score 73.6\%), allowing us to do a descriptive \emph{explanation} (rather than \emph{prediction}) of the code-opaque features. To this end, we considered their relative feature importance ranking. This indicated that performance regressions tend to be associated with larger diffs from either very junior or very senior authors (convex curve) in certain rarely touched sub parts of the system.

\section{Bag of words model}

\label{bm25:vec}
As results of code-opaque models were not satisfactory for prediction (see \Cref{sec:code-results}), we moved to models that could ingest the \emph{actual code changes}. The initial goal was to understand how well relatively simple traditional ML approaches would perform, particularly since they had shown to be effective for similar problems in industry deployments~\cite{murali2021industry}. To this aim, we used a Bag of Words (BoW) representation of code~\cite{voorhees1999natural}.

\subsection{Method}

A BoW model is a simplifying representation of a textual document originating from natural language processing, which has recently been adopted in Software Engineering~\cite{huo2016learning}. In this model, a piece of code is represented as the multiset of its words, disregarding structural information from the code and even statement order, but keeping multiplicity.

In our BoW model, we input the textual code changes from a diff (see \Cref{sec:diffbert,sec:codecontext}). After removing all the numbers and symbols, we determine word boundaries through whitespaces and other syntactical stop tokens such as an opening parenthesis. From each word, we create sub-words based on snake-case and camel-case. Once all the sub-words have been acquired, we convert them to lower-case. This way, {\tt newFunctionCall()} is determined to be its own word through the preceding space and the succeeding opening parenthesis and gets expanded into the sub-words {\tt new}, {\tt function}, {\tt call}. This fine-grained strategy, which builds the vocabulary based on the train set, reduces dimesionality and thus makes the occurrence of out-of-vocabulary words in the test set rare by construction. Due to this, we simply ignore missing sub-words in the test set.

For example, the code change in Listing~\ref{lst:diff} were to be translated into its BoW representation in Listing~\ref{lst:diff_bow}.  Note in particular how code context is not prefaced with either a {\tt +} or {\tt -} sign, and how they otherwise prefix each token. \Cref{tab:freq_tab} shows how the sub-words are collected and how the input is finally vectorized and then fed as-such to the learner.

We further process the occurrence frequency  of each such sub-word with the Okapi BM25 vectorizer~\cite{robertson1995okapi}. We did this because BM25 is a state-of-the-art TF-IDF-like retrieval function~\cite{whissell2011improving}, and goes beyond simply rewarding term frequency and penalizing document frequency by accounting for document length and term frequency saturation.

\begin{listing}[h]
\begin{minted}
[
frame=lines,
framesep=2mm,
baselinestretch=1.2,
bgcolor=LightGray,
fontsize=\footnotesize,
escapeinside=@@,
startinline
]{text}
A line of code context
+ import static com.example.animport;
A line of code context

A line of code context
+ // code comment
- oldFunctionCall()
+ newFunctionCall()
A line of code context
\end{minted}
\caption{\texttt{diff} representation of a code change}
\vspace{-0.4cm}
\label{lst:diff}
\end{listing}

\begin{listing}[h]
\begin{minted}
[
frame=lines,
framesep=2mm,
baselinestretch=1.2,
bgcolor=LightGray,
fontsize=\footnotesize,
escapeinside=@@,
startinline
]{text}
a line of code context
+import +static +com +example +animport
a line of code context
a line of code context
-old -function -call
+new +function +call
a line of code context
\end{minted}
\caption{BoW representation of the code change in Listing~\ref{lst:diff}}
\label{lst:diff_bow}
\end{listing}

\begin{table}[tb]
\caption{BoW pre-vectorization example.}
\label{tab:freq_tab}

\resizebox{\linewidth}{!}{

\begin{tabular}{lr|lr|lr|lr}
\toprule
\multicolumn{1}{l}{\textbf{Sub-Word}} & \multicolumn{1}{l}{\textbf{\#}} & \multicolumn{1}{l}{\textbf{Sub-Word}} & \multicolumn{1}{l}{\textbf{\#}} & \multicolumn{1}{l}{\textbf{Sub-Word}} & \multicolumn{1}{l}{\textbf{\#}} & \textbf{Sub-Word}             & \textbf{\#}    \\
\midrule
\tt{a}                                     & 4                                      & \tt{+import}                               & 1                                      & -old                                  & 1                                      & \tt{+new}      & 1 \\
\tt{line}                                  & 4                                      & \tt{+static}                               & 1                                      & \tt{-function}                             & 1                                      & \tt{+function} & 1 \\
\tt{of}                                    & 4                                      & \tt{+com}                                  & 1                                      & \tt{-call}                                 & 1                                      & \tt{+call}                         & 1 \\
\tt{code}                                  & 4                                      & \tt{+example}                              & 1                                      &                                       &                                        &                               &                       \\
\tt{context}                               & 4                                      & \tt{+animport}                             & 1                                      &                                       &                                        &                               &                      \\
\bottomrule
\end{tabular}
}
\end{table}

\subsection{Results}
\label{sec:bowresults}

Results for the BoW model in \Cref{fig:bow_results} depict a significant improvement over the code-opaque models, achieving an AUC of 84\% and an average precision of 14\%. Moreover, we can observe that adding more code context helps increase prediction accuracy, but that we likely hit marginal rate of returns at around 7 lines of context. However, these results are not good enough yet for predicting regressions in a production setting, as the confusion matrix in \Cref{tab:confusion} shows.

\begin{table}[tb]
\centering

\caption{Confusion matrix for BoW model.}
\label{tab:confusion}

\resizebox{0.8\linewidth}{!}{

\begin{tabular}{lllll}
\toprule
& \textbf{Precision} & \textbf{Recall} & \textbf{f1-score} & \textbf{Support}
\\ \midrule
False & 0.99 & 1.0 & 0.99 & 14,285 \\
True & 0.75 & 0.01 & 0.03 & 200 \\
\bottomrule
\end{tabular}
}
\end{table}

This model only finds 1\% of all regressions (true cases). However, it is correct for 75\% of the time when it does predict a regression. Whenever it predicts a change to be benign, it almost always is (99\% of cases). Conversely, when we tune the prediction threshold to have a 75\% recall for regressions, precision decreases to 4\%, and recall for non-regressions to 76\%. In the test set, 150 of 200 regressions will be caught in this scenario, while mislabeling 3,478 non-regressions (out of a total of 14,285 non-regressions).

\section{CodeBERT-based Model}
\label{sec:codebert}
Architectural shortcomings of the BoW model are obvious: its dimensionality and sparseness poses practical challenges, as BoW models are ultimately only a crude representation of code.

Particularly problematic for SuperPerforator, the BoW vector representation of the following two code changes would be identical:
\begin{listing}[h]
\begin{minted}
[
frame=lines,
framesep=2mm,
baselinestretch=1.2,
bgcolor=LightGray,
fontsize=\footnotesize,
escapeinside=@@,
startinline
]{python}

+ a = not_expensive_function(user)
\end{minted}
\end{listing}

This first code change would likely be a non-regression, whereas an easy-to-miss change could make it risky:


\begin{listing}[h]
\begin{minted}
[
frame=lines,
framesep=2mm,
baselinestretch=1.2,
bgcolor=LightGray,
fontsize=\footnotesize,
escapeinside=@@,
startinline
]{python}

+ a = not expensive_function(user)

\end{minted}
\end{listing}

We surmised that by incorporating such information, we achieve another step change in performance, similar to the move from code-opaque to code-transparent models. Models based on the CodeBERT architecture allow us to do this and have generally beaten the state-of-the art results for a plethora of Software Engineering related tasks~\cite{mashhadi2021applying,zhang2021improving,zhou2021assessing}. As such, a natural thought was to use CodeBERT to model our task of regression prediction.

\subsection{Method}
We treat our task of regression prediction as a binary text classification problem to CodeBERT.

To this aim, we used the CodeBERT pre-trained tokenzier and model from Huggingface~\cite{huggingface} and fine-tuned it with the training set from \Cref{sec:thedata}. Because CodeBERT does not come with functionality to handle the high imbalance present in our data set, we added a cross entropy layer on top of the neural net to assign class weights reciprocal to their respective occurrence frequency during training. Moreover, we used the evaluation $f1$ score to guide training and model selection. As the harmonic mean of precision and recall, using $f1$ is a best practice when dealing with highly imbalanced data~\cite{cruz2016tackling}.

\subsection{Results}

\Cref{fig:bow_results} (CodeBERT's sub plot omitted; it did not lift from the x-axis, similar to the first sub-plot) and \Cref{tab:bestmodel} show that even with these improvements and after training for a total of 15 epochs or 9 hours, results for the CodeBERT models are significantly worse than the BoW model, and comparable to the code-opaque feature model. Due to the poor results, we omit graphs for multiple context lengths. These results are in spite of a decreasing learn loss in the model, selecting the best model based on performance of the evaluation data set, training for enough epochs, appropriate setting of the learning rate and respecting other DL best practices.

Closer inspection of the generate predictions shows a tendency of the model to pickup predicting only one class, even though tokenization of the input code changes looks reasonable, with some input tokens being out of vocabulary for CodeBERT. Moreover, we also observe that a substantial amount of the input diffs is truncated due to CodeBERT's maximum document length limit. Finally, CodeBERT was not designed to handle diffs (\emph{i.e.,} code patches formatted by the Unix {\tt diff} utility) , but rather direct source code. Therefore, it is unclear how well it copes with this new, partly repetitive input structure and whether it is able to learn to extend the attention of special tokens such as {\tt +} or {\tt -} to the rest of the line. To sum up, in addition to the domain shift by now operating on a language CodeBERT was not trained for, there is another domain shift by having formatted diffs as input to the model.

\section{SuperPerforator}

Given our experiences from CodeBERT in \Cref{sec:codebert}, it seemed reasonable to assume that its performance could be greatly improved by using the related, but better apt DiffBERT.

\subsection{Method}
\label{sec:method}

Our initial experience when training DiffBERT hinted at the fact that this might be a challenging problem: While learn loss steadily declined over 10 epochs (and only plateaued around 13 epochs), evaluation loss monotonically increased, signifying that the model started to over-fit on the train data from the start. Moreover, it also learned to predict only the majority class, similarly to the CodeBERT model.

To this end, we first created a SuperPerforator-bespoke pre-training that eschews DiffBERT's non-code features. We pre-trained DiffBERT on 10 million code changes from the internal code-base. The model architecture utilized by DiffBERT is based on the Longformer~\cite{beltagy2020longformer} with 12 encoder layers, 512 token attention window, 768 hidden size, 50,005 vocabulary size (5 special tokens, \emph{e.g.,} for added or removed lines, new lines, ...), and a sequence length of 2,048. The pre-training task is a masked language modeling task with a masking probability of 0.15. Most of these hyper-parameter values are the ones used in the original Longformer and BERT models. It was run for 5 epochs on 192 GPUs, each with a batch size of 5 and a total wall runtime of 25 hours. The training log loss for pre-trained DiffBERT is 0.12 on 10 million samples; the evaluation loss is 0.06 on 200,000 samples.

Secondly, we included a Cross Entropy Loss layer and evaluation using the $f1$ score during training (the same as described for CodeBERT, see \Cref{sec:codebert}).

After pre-training, we fine-tuned DiffBERT for the performance regression prediction task with the binarized data from \Cref{sec:thedata} for a maximum of 10 epochs with early stopping. Patience value 2 is used for early stopping where fine-tuning is terminated if no improvement on the validation data $f1$ score is observed for two consecutive epochs.

\subsection{Results}

As opposed to the CodeBERT model, the improvements in DiffBERT made the model sensitive to the input data, allowing it to perform sensible predictions based on variances in the input embeddings.

The resulting fine-tuned DiffBERT model contained a total of 147 million trainable parameters; its fine-tuning took roughly 19 hours on an 8-GPU machine. Offline results in \Cref{tab:bestmodel} demonstrate almost on-par performance with the best BoW model, and as such, a textual description of its prediction performance would be similar to the one in \Cref{sec:bowresults}. Similar to the BoW model. \Cref{fig:bow_results} shows an albeit non-monotonic trend to improvement given longer contexts.

\begin{table}[tb]
\caption{Comparison of best-per-class model in offline setting.}
\label{tab:bestmodel}

\centering
\begin{tabular}{lrr}
\toprule
\textbf{Model} & \textbf{$f1$ (minority class)} & \textbf{Training time} \\ \midrule
Code-opaque  & 0.01 & $<$ 5m\\
BoW & 0.30 & 15m\\
CodeBERT & 0.02 & 9 hrs\\
DiffBERT  & 0.28 & pre-training: 25 hrs\\
& & fine-tuning: 19 hrs\\
\bottomrule
\end{tabular}
\vspace{-0.3cm}
\end{table}

\subsection{Transfer learning for \AdsManager}
\label{sec:tl}
\begin{figure*}[tb]
\centering
\includegraphics[width=1.6\columnwidth]{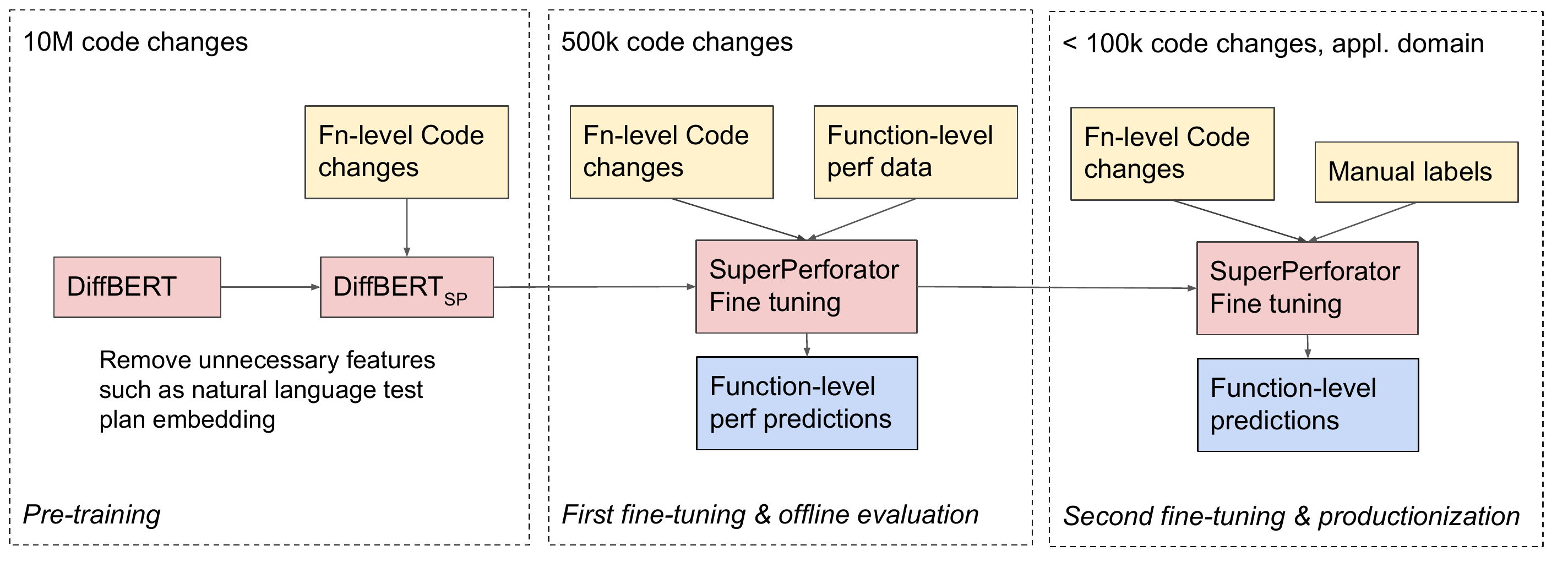}
\vspace{-0.3cm}
\caption{SuperPerforator's architecture and how it evolved from plain DiffBERT.}
\label{fig:superperforator}
\end{figure*}

\begin{table}[tb]
\caption{Online comparison of SuperPerforator and BoW.}
\label{tab:comp-sp-bow}

\centering
\begin{tabular}{lrr}
\toprule
\textbf{Property} & \textbf{BoW} & \textbf{SuperPerforator} \\ \midrule
Transfer learning  & \ding{55} & \ding{51}\\
Few Shot learning & \ding{55} & \ding{51}\\
Iterative training & \ding{55} & \ding{51}\\
Best Offline Mean Average Precision & 0.2 & 0.19 \\
\bottomrule
\end{tabular}
\vspace{-0.5cm}
\end{table}

Our results highlight that both BoW and DiffBERT-based models might be equally well-suited suited to filter out benign changes with high certainty, due to their similar best average precision, i.e. the average mean precision between regression and non-regression cases for the best-performing model of each variant (BoW and SuperPerforator). However, the DiffBERT-based SuperPerforator model has additional advantages highlighted in \Cref{tab:comp-sp-bow}: it can encode diffs from any language into a dense representation, leverage previous training to do few-shot and transfer learning, and be trained iteratively, giving a bias towards recently added data, a property that is important given the established temporal component of performance predictions (see \cref{sec:chrono}).

To productionize SuperPerforator for \AdsManager, we applied a secondary fine-tuning process to SuperPerforator's base model to train it on its final use domain. \AdsManager supplied us with a manually labeled set of regressions for its part of the code base. SuperPerforator can then be incrementally refined by fine-tuning it whenever new \AdsManager training data is available on top of the previous fine-tuning result. This also leads to newer data having more weight and a much shorter training procedure.
\Cref{fig:superperforator} depicts SuperPerforator's final architecture, which we can break into a (1) pre-training, (2) first fine-tuning, and (3) second fine-tuning and productionization phase.

\AdsManager data differs in two cardinal aspects from that of the base model: (1) it contains JavaScript frontend code and (2) the imbalance ratio is even higher, with only 11 regressions for all code changes in the six months from December 2020 to June 2021.

For \AdsManager, we deployed SuperPerforator in filtering mode as the first of a cascade of tools to flag regressing diffs before they make it to production, tuning our model for recall: based on historic data from December 2020 to June 2021, we chose a classification threshold that achieves perfect recall. Applying this same threshold for the experimentation period from June to December 2021 confirmed the results, again missing no regression. Overall, pre-filtering via SuperPerforator reduces the space of diffs which \AdsManager needs to run downstream regression detection tools on by 43\% (backend) and 27\% (frontend) of diffs respectively. These results substantiate an up to 45x improvement over a random baseline, which only used the prior probability of a regression being present as its input. It also marks a ~10x online improvement over the BoW model in the same setting, despite similar offline performance (due to confidentiality, we can not share actual production performance figures).

\subsection{Counterfactual explanations}

To help us validate and understand SuperPerforator's DL model, we generate explanations that can highlight parts of the code change that the model deems significant for its prediction.
Explanations benefit two distinct sets of users: (1) They can aid debugging for the developers, by helping them understand whether a model has picked up on spurious correlations.
(2) They can also benefit the end users by helping them reconstruct the rationale of the prediction and point to particular parts of the code change that the model thinks is the reason for causing the performance regression.

There are several explanation methodologies for DL, including methods based on attention~\cite{attention1, attention2}, perturbation (e.g., LIME~\cite{LIME}, SHAP~\cite{SHAP}), or on the model's gradient~\cite{ig}.
All methods assign a numerical approximation to each token in the input that represents a notion of importance towards a particular prediction.

We decided to implement counterfactual explanations~\cite{cito:22}, a perturbation-based approach that computes minimally different alternate ``worlds" (\emph{i.e.,} inputs to the model) that lead the model to make a different prediction.
For SuperPerforator, we pose the question: what is the smallest possible alteration to the diff for the model to change its prediction? This points us to parts of the input that were important for the model to arrive at its prediction.

\subsubsection{Method}

Computing counterfactuals can be posed as a search problem where the input to the model is perturbed until the prediction ``flips" (\emph{i.e.}, changes to determining the altered code change is not a performance regression any longer).
This search is generally intractable~\cite{cf_survey}.
To compute sensible counterfactuals with reasonable efficiency, we introduce a set of constraints on the search space known as plausibility and sparsity.
Plausibility means that the resulting counterfactual should be conceivably part of the original input distribution of the model.
Sparsity implies a minimality constraint on the introduced perturbations on the original input that generate the counterfactual.
This means that we should perturb the input as little as is required to flip the decision.
To generate plausible counterfactuals, we use masked language modeling as a perturbation operator (similar to counterfactual text generation~\cite{madaan:21}), which is available through via DiffBERT's API. We mask a set of token(s) in the original input and ask the model to generate in-distribution alternatives.
We then systematically build a search space by repeatedly applying these perturbations through a greedy search procedure~\cite{nemhauser:78, martens:14}, guided by the internal model score of each counterfactual candidate.
For SuperPerforator, we limit perturbations in the code change input to function and method invocations and import statements.
\Cref{lst:counterfactual} shows an example of a simple counterfactual explanation.
It says, \emph{if you called \texttt{call\_cheap\_function} instead of \texttt{call\_medium\_expensive\_function}, SuperPerforator would no longer deem the code change to induce a regression}.

\begin{listing}[tb]
\begin{minted}
[
frame=lines,
framesep=2mm,
baselinestretch=1.2,
bgcolor=LightGray,
fontsize=\footnotesize,
escapeinside=@@,
startinline
]{python}
for i in (1, all_users):
    print(i)
+   @\colorbox{red}{call\_medium\_expensive\_function(i)}@
    @\colorbox{green}{call\_cheap\_function(i)}@
\end{minted}
\caption{Counterfactual explanation for SuperPerforator's prediction that the code change  will cause a regression.}
\vspace{-0.7cm}
\label{lst:counterfactual}
\end{listing}

\subsubsection{Method}

We performed a qualitative exploratory study with two software developers at \Meta by showing them counterfactual explanations for SuperPerforator predictions for 20 randomly sampled code changes from our test set.
We also assessed whether the software developers found the explanation useful \emph{and} whether they think, based on seeing the explanation, that the prediction is a true- or false-positive.
The two study participants were able to determine the correct label in 85\% (17/20) of cases, and it equally useful. We noticed a distinction in how participants came to the conclusion on how to interpret their explanation. In true-positive instances, participants noted that the explanation guided them to parts of the code that were aligned with their mental model. More specifically, the explanation reinforced their hypothesis that had been induced through the prediction. One participant did note that while this line of reasoning seems sound, it could be prone to confirmation bias.
In false-positive instances, participants noted that the strongest signal not to trust the prediction was the level of unreasonableness of the explanation. If the
explanation pointed to irrelevant parts of the input or introduced an irrational perturbation, it was a sign that the prediction could probably not be trusted.

Overall, our qualitative analysis showed that the counterfactual explanations provided  developers with intuition and
confidence in understanding exactly where the model is picking up its signal. Whether that signal was reasonable or not helped them
decide to trust or discard the prediction.




\section{Discussion}

In this paper, we performed a comparative study of four different models to predict the presence of performance regressions before they land in production. We learned that prediction of performance regressions is an inherently hard problem, starting with a large imbalance in the data sets, and an important temporal aspect to it that renders a standard random train-test split ineffective. The hardness of the problem is surprising given how well sequential DL models, even plain CodeBERT, have performed for other, seemingly related tasks, \emph{e.g.,} defect prediction~\cite{pan2021empirical}.

Results on the code-opaque and BoW models demonstrate that ingesting the actual source code changes is paramount for successful performance prediction. Providing more code context generally helped, and reached diminishing returns at around 7 lines. SuperPerforator showed slightly paradoxical behavior, which was not a limitation of the underlying maximum document length and requires further research.

Our experiments generally also show that more complex model families exhibit step-change increases in prediction performance over the most simple code-opaque model. Performance of simple code-opaque models was unacceptable for practical uses. BoW models performed significantly better. Despite starting at a disadvantage, through adaptation and customization, we were able to achieve on-par results with the bespoke Transformer-based model architecture SuperPerforator. Lastly, SuperPerforator showed excellent transfer learning capabilities in production use, which make it far superior to the BoW model. This is because DiffBERT (1) uses a sub-word tokenizer, which means that there is substantial overlap between code changes in various programming languages (also to help deal with the out of vocabulary problem, \cite{pinter2017mimicking}), (2) has been pre-trained on a wide number of different languages, apparently enabling it to work on a language not seen during fine-tuning (but very similar to one that is known), (3) can be incrementally refined by fine-tuning it week-over-week, starting with the previous model as a base for the weights.

While the DiffBERT-based SuperPerforator model hinges onto semantic features, it is less apt at embedding structure. Particularly problematic for SuperPerforator, the representation of the following two code changes would be near-identical, as a manual comparison of their embedding in a two-dimensional representation in vector space revealed:

\begin{listing}[h]
\begin{minted}
[
frame=lines,
framesep=2mm,
baselinestretch=1.2,
bgcolor=LightGray,
fontsize=\footnotesize,
escapeinside=@@,
startinline
]{python}

+ for i in (1, all_users):
+   print_if_debug(i)
+ call_medium_expensive_function(i)

\end{minted}
\vspace{-0.4cm}

\end{listing}

This first code change would likely be a non-regression, whereas a small structural change could easily convert it to a highly suspicious code change:
\begin{listing}[h]
\begin{minted}
[
frame=lines,
framesep=2mm,
baselinestretch=1.2,
bgcolor=LightGray,
fontsize=\footnotesize,
escapeinside=@@,
startinline
]{python}

+ for i in (1, all_users):
+   print_if_debug(i)
+   call_medium_expensive_function(i)

\end{minted}
\vspace{-0.8cm}
\end{listing}

We surmise that by incorporating structure into the model, we might achieve another step change in performance. Models such as GraphCodeBERT could accomplish this, but were out-of scope for this principal investigation into performance regressions~\cite{guo2020graphcodebert}. Moreover, the additional benefit of adding graph information has lately been put into question~\cite{karmakar2021pre}. This also raises the question of whether the performance prediction task has an inherent property that makes it ill-suited for Transformer-based architectures, another avenue for future research. Lastly, it could be that the task is so complex that it requires the use of super-large models with billions of parameters, an order of magnitude more than SuperPerofrator's 147 million, following recent advances in big code~\cite{brown2020language}.

\section{Threats to Validity}

\subsection{Internal validity}
Code opaque models work at the diff-level (for they use diff-level features such as the author of the code), while the other models studied here work at a function-change level, which can potentially encompass a subset of changes from multiple diffs. Naturally, code context cannot be varied when there is no code. This makes a completely identical comparison between these models impossible. However, we have tried to bridge the gap between the two by only using changes that arose from one diff and that only contain changes to one function for the code-opaque models. We ended up with 94\% of the original data for our code-opaque level experiments. There is no domain-shift in the two distributions.

Moreover, a fair comparison between DL and non-DL techniques is in general complex. Usually, different methods benefit from more or less train data in different ways. To counter for the observed high chronological component in predicting performance regressions, we established a priori fixed train-test sets and used them for all algorithms. In return, DL models were allowed to train until they reached saturation by achieving their respective early stopping criterion.

We put different amount of effort into each of the models. This could lead to us under-reporting the theoretically possible performance of the models we put less work into (\emph{e.g.}, by missing opportunities in feature engineering). To mitigate this effect, we only stopped model refinement for a given architecture when we observed diminishing returns, which made it plausible that no significant improvement was to be expected. In the end, performance gaps between models are so vast that even if a (however unlikely) doubling of performance in the inferior model was possible, the superior model would still greatly outperform it.

\subsection{External validity}
We developed our approach specifically for use at \Meta, using internal tooling such as \FBDetect and DiffBERT. One challenge in particular for academic researchers might be the large amount of in-production performance measurements that is required. This could be enabled if large open-source organizations shared similar data in the future. As opposed to many other DL models, we emphasize that hardware availability or training duration should pose no hindrance in our case.

\section{Conclusion \& Future Work}
In conclusion, this work shows that forecasting function performance regressions is a challenging ML task. Light-weight solutions incorporating only code-opaque features showed some explanatory power, but were not enough to make precise predictions about a given piece of code. More advanced models incorporating the actual changed code outperformed such simpler attempts by an order of magnitude.

We then directed our attention toward DL-based models, with the idea that they would even better understand the intricate connections between code patterns, function invocations, code syntax, and their impact on performance. Results on a vanilla CodeBERT show that DL-based models initially disappointed, and only came on-par with naïve BoW-based models after pre-training and customization. A large number of accurate performance measurements (in the $10^5$) and pre-processing was necessary just to get the DL-based models to train and generalize to unforeseen data. An exploration into which parts of a code change the model deemed important based on counterfactual explanations delivered plausible results, thereby not only validating the generated black box SuperPerforator model but also promising to help software developers interpret its performance predictions in the future.

In this paper, we have shown that that fully automated performance prediction of software is hard, but possible. The deployment of SuperPerforator at \Meta shows that such models can be superior to simpler models in an online setting despite achieving similar offline performance because they enable transfer-learning, few shot learning, and iterative refinement with recency.  The area of performance prediction using ML is still in its infancy, though: Are performance regressions outside of \Meta equally hard to predict? Why do Transformer-based models perform so poorly out-of-the-box, when they have redefined automation in Software Engineering elsewhere? Is the Transformer architecture inherently unfit to the task of performance classification? Would models that incorporate more code structure, such as GraphCodeBERT, help? Finally, as SuperPerforator presents a purely functional view on code changes, what is the role of non-code changes (\emph{e.g.,} configuration or hardware) on code performance and how can they be modeled?

\newpage

\bibliographystyle{IEEEtran}
\tiny
\bibliography{IEEEabrv,paper}
\end{document}